# Developing spatiotemporal solitons in step-index multimode fibers

Yao Wang, Shuo Yan, Jianqiu Xu & Yulong Tang[*]

**Spatiotemporal solitons (STSs) are localized solitary waves in both space and time that involve complex linear and nonlinear processes. Optical STSs have been observed in various media, but they are difficult to be realized in multimode fibers due to their large modal dispersion. Here, we report STS mode-locking and spatiotemporal nonlinear dynamics in a step-index multimode fiber mediated by gain. Gain competition and energy redistribution among different spatial modes help nonlinearity effectively cancel both temporal splitting and spatial diffracting of the entire pulse (multicomponent soliton) and thereby maintain its shape during propagation. Optical STSs in multimode fibers therefore open a new way to clarify the fundamental science of spatiotemporal solitary waves and also imply important applications in multi-channel communications, optical switches and mode-area power scaling of fiber laser pulses.**

Key Laboratory for Laser Plasmas (Ministry of Education), Department of Physics and Astronomy, and IFSA Collaborative Innovation Center, Shanghai Jiao Tong University, Shanghai 200240, China. Correspondence and requests for materials should be addressed to Y. T. (Email: yulong@sjtu.edu.cn).

Solitary waves (solitons) are among the most mysterious natural phenomena and have attracted scientists' attention for centuries. They can occur in various circumstances, such as liquids, plasmas, condensed matter and fibers[1-4], and are usually explained by interplay between linear and nonlinear processes. However, the recent observed three-dimensional optical solitons, also named spatiotemporal solitons (STSs) or light bullets, have stimulated new prospects in both fundamental sciences and applications[5-7]. Optical STSs are light pulses that propagate in media with neither dispersive broadening nor diffractive spreading, which only occur when the Kerr nonlinearity the pulses experience can exactly balance both temporal dispersion and spatial diffraction.

Solitons generated in single mode fibers are the most successfully explored optical solitary waves, and have brought great impact on telecommunications. On the contrary, the formation and evolution of solitons in multimode fibers, which incorporate the spatial degree of freedom, are still seldom touched upon. Though many early works[8,9] have proposed and thereafter some theoretical reports[10,11] have confirmed that multimode fibers can support STSs (also named multimode solitons), only very recently have their formation and propagation dynamics been experimentally explored[12,13]. Soliton trapping and energy transfer within multimode solitons have also been theoretically observed[14]. However, all these works only dealt with the issues in passive and graded-index fibers. As for multimode gain fibers, the addition of gain will strengthen coupling between the spatial and temporal degrees of freedom (beam size, pulse duration, etc.), thus bringing new balance between linear and nonlinear processes, and will trigger unique soliton dynamics. In addition, the large modal dispersion of step-index multimode fibers is beneficial for compensating large chromatic dispersion, facilitating generation of broadband multimode solitons. Therefore, step-index multimode gain fibers provide an excellent platform to explore the fundamental mechanisms and abundant nonlinear dynamics of multi-dimensional optical solitary waves.

Apart from widespread scientific interest, STSs in multimode fibers can also find a number of important applications, such as multiple-channel communications[15], dispersive multiplexing[16], space-division multiplexing[17], imaging[18], and cell manipulation[19], etc. In addition, multimode solitons have much larger mode area than their single-mode counterparts, which is very beneficial for mitigating detrimental nonlinear effects, and thereby can accommodate high-energy laser pulse generation and transmission.

Here, we develop STS mode-locking in 2 μm (anomalous dispersion) step-index multimode thulium fibers for the first time. Through carefully tuning the coupling among transverse modes and exploiting modal energy transfer, STS mode-locking is experimentally achieved, during which both the temporal and spatial shapes (pulse duration and beam size) of the multimode soliton can be self-adjusted and consequently become localized. Phenomena such as soliton attraction, spectral self-arrangement and energy concentration to the fundamental mode are investigated, revealing underlying mechanisms and unique nonlinear dynamics. Strong spatiotemporal coupling is found to be beneficial for both stabilizing the multimode solitons and scaling their pulse energy. As a preliminary example, our multimode soliton fiber laser achieves >10 W of average power and >500 nJ of pulse energy in the picosecond regime.

**Results**

In order to accommodate multiple transverse modes and, at the same time, not to cause too much material dispersion, 3.5 m length of highly doped thulium fiber with 25/400 μm core/clad is adopted as the gain material. The comparatively short length of gain fiber can also efficiently decouple gain from group velocity dispersion (GVD) and nonlinear phase shift[20]. Based on the lasing wavelength of ~2 μm, the normalized frequency of the fiber is V~5.88, thus this fiber can support multiple transverse modes. Experimental setup for the fiber system with a linear cavity is schematically shown in Fig. 1. To achieve mode-locking of multiple spatial modes, efficient nonlinear coupling between the fundamental mode and higher-order modes is needed, as schematically shown in Fig. 1a. Certainly, such coupling mainly happens inside the fiber core through light-material interaction. It is this kind of nonlinear modal coupling that leads to the formation of STSs and related complex soliton dynamics. Figure 1c schematically shows the simulated intensity profile of several lowest order modes and their overlapped profile, while Fig. 1d schematically shows the temporal pulse shape of three individual modes and the total pulse shape under three different overlapping cases.

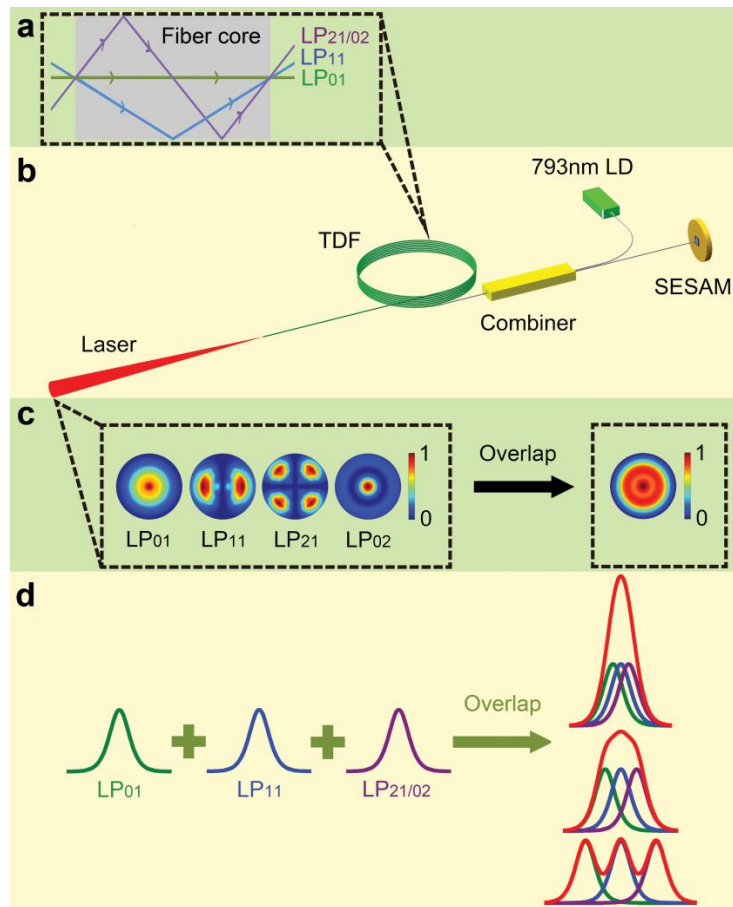

**Figure 1. (a) schematically shows the coupling between the fundamental mode and higher order modes in the fiber core; (b) is the experimental setup with related components of the mode-locked multimode fiber laser; (c) schematically shows the simulated intensity profile of several spatial modes and the final overlapped one in this multimode fiber laser; (d) schematically indicates the temporal pulse profile of individual solitons and the overlapped multimode soliton.** In (d), the red lines denote the overlapped pulse shapes. TDF: thulium doped fiber; LD: laser diode; SESAM: semiconductor saturable absorber mirror; HOM: higher order mode; FM: fundamental mode.

In experiment, increasing the 793 nm pump power first stimulates the laser to the continuous-wave (CW) lasing with a threshold of ~7 W. Through slightly increasing the pump and carefully tuning the cavity, the fiber laser transits to one stable CW multimode mode-locking region (region I in Fig. 2a). At a given power level, the multimode soliton shows variable output beam profile (Supplementary Fig. S1), which conforms to the primary experimental signatures of multimode mode-locking [21]. In the mode-locking state, no Q-switching or Q-switched mode-locking is observed due to the large output coupling ratio[22]. Stable operation of multimode solitons, involving both the temporal and spatial degrees of freedom, requires simultaneously locking not only the longitudinal modes but also the multiple transverse modes, meaning a big step forward from conventional single-mode solitons.

This table mode-locking can be sustained within a certain pump power range (~7 W pump), and further increasing the pump to over 15 W brings the state to be unstable. At this time, slightly/carefully tilting the SESAM and/or varying the gap between the fiber end and the SESAM (fulfill right dispersion compensation for different modes) returns the pulsing to stable mode-locking, and the laser goes into the next stable region (region II). This tuning-back-to-stable process repeats and stable mode-locking operation can be achieved at over 10 W output power levels. Therefore, several discrete stable multimode-locking regions (from I to V) form across the whole pump range. Detailed output characteristics of the mode-locked multimode fiber laser are indicated in Fig. 2. The maximum output power and pulse energy are 10.6 W and 550 nJ, respectively, clearly giving record performance in 2-µm mode-locked fiber oscillators. The pulse energy is by far larger than that of the 2-µm single-mode fiber solitons[23-27]. This thus opens an efficient way to achieve high power/energy laser pulses directly from fiber lasers. The output power is only limited by damaging the mode-locking element, the SESAM used. Higher output power is expected if the beam size impinging on the SESAM is intentionally expanded.

Figure 2b displays the pulse train of the mode-locked fiber laser at the 10-W output level, showing a repetition rate of ~20 MHz (corresponding to the round-trip time of ~50 ns). The radio-frequency (RF) spectrum measured with 1 kHz resolution has a >65 dB signal-to-noise ratio (Fig. 2c), showing high stability of the multimode-locking. The stability is further confirmed by the RF spectrum measured under a 320 MHz frequency span (Fig. 2d).

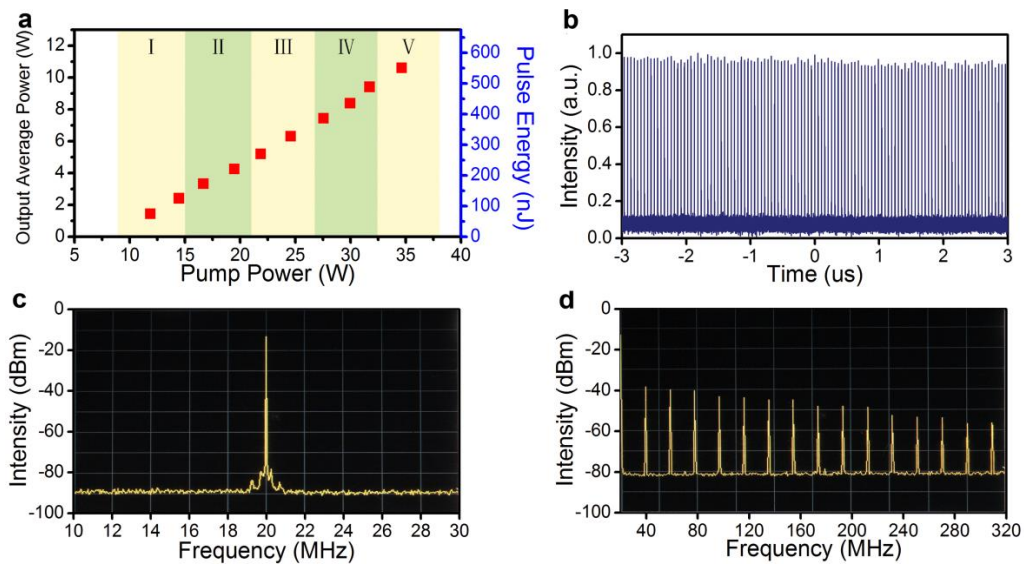

**Figure 2. Output characteristics of the mode-locked multimode fiber laser (a), and the output pulse train (b), RF spectra in 30 MHz span (c) and 320 MHz span (d) measured at the 10 W level.** In (a), column areas with different colors show discrete stable mode-locking regions**.**

Laser spectra measured under different power levels (together with the near field laser beam profiles) are shown in Fig. 3. Detailed investigation of the spectral evolution can uncover the underlying mechanisms and affluent nonlinear dynamics in multimode solitons. Under low power levels, the unique three separated spectral peaks (Fig. 3a) clearly mean that the multimode solitons (temporal shape, see Fig. 4a) include three different transverse modes. This has been confirmed by the measurement with the variable round aperture method (Supplementary Fig. S2), where the spectral peak number will reduce (from three to two, and then to one) when the light-pass aperture is diminished. The longest-wavelength (2024 nm) peak is ascribed to the fundamental mode (LP01), while the second- (2015 nm) and the shortest-wavelength (1994 nm) peaks to the high-order modes of LP11 and LP21/02, respectively. Different spatial modes possessing different center wavelengths are indispensible for the formation of stable multimode soliton. Only under this condition can the material dispersion (anomalous at 2 μm) effectively balance the large modal dispersion, and consequently these three transverse modes sychronize their propagating (overlap temporally), and thereby coalesce and act as a multicomponent soliton during evolution in the fiber. Each discrete spectrum has its own Kelly sidebands (refer to Supplementary Fig. S3 for an amplified spectrum in linear scale), showing mode-locking occurs in the anomalous dispersion region[28]. Figure 3a shows that the highest order mode (corresponds to the shortest wavelength) dominates, but this can be easily changed by tilting the SESAM to introduce different loss to different spatial modes (compare this situation with Supplementary Fig. S2).

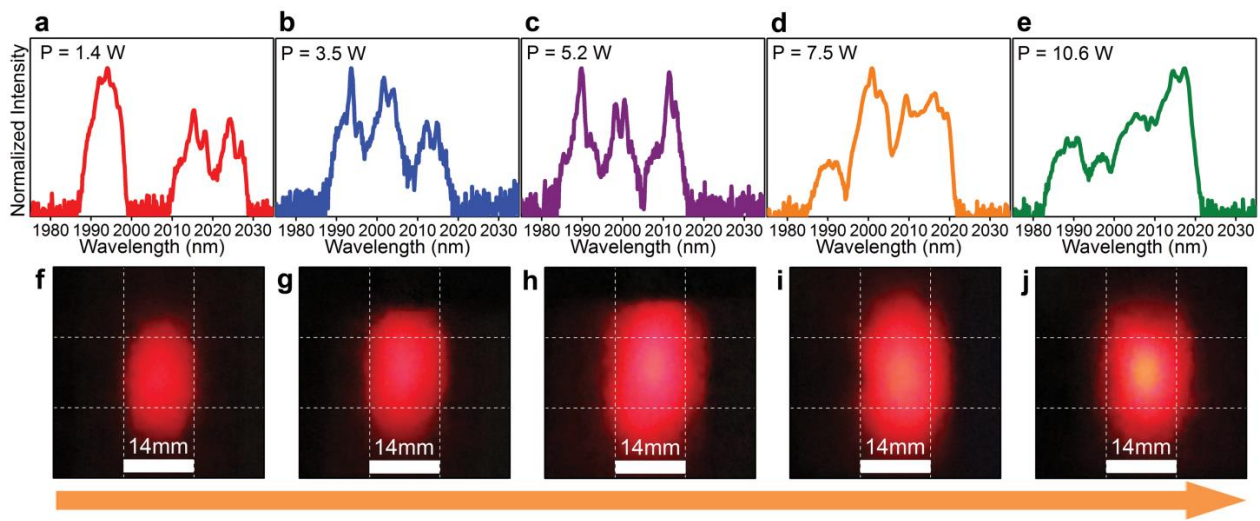

**Figure 3. Laser spectra (a-e) and laser beam intensity profiles (f-j) of the mode-locked multimode fiber laser under different output power levels.** All the spectra are normalized to their highest peak in the logarithmic scale. Each beam profile in the second row corresponds

to the spectrum on the top of it in the first row. The beam profile is measured at 66 mm away from the output fiber facet, and the bottom arrow shows the power-increasing direction.

When the output power is increased to 3.5 W, the fundamental (LP01) and the first higher-order (LP11) modes show spectral blue-shift while the highest-order mode (LP21/02) indicates spectral red-shift, thus leading to speed up of low-order modes and slow down of high-order modes, improving synchronization of the whole pulse. Spectral converging is a manifestation of intra-soliton attraction among different modes in multimode fibers due to Kerr nonlinearity (XPM)[29]. Nonlinear spatiotemporal coupling of different modes gives rise to the attractive force between different modes, strengthens their temporal overlap, and thus narrows the total pulse duration (Fig. 4a). At this time, the energy of both the fundamental and higher-order modes increases, leading to an increased beam size (see also Fig. 4c).

Increasing the output power further to 5.2 W, the spectrum shows some instability due to strong inter spatial-mode interaction and energy exchange. Other high-order modes beyond LP21/02 (with varying center wavelengths) sometimes appear (Supplementary Fig. S4c). Consequently, existence of more modes together with instable soliton attraction reduces the temporal overlap and leads to stretching of the pulse to ~42.62 ps (Fig. 4a). At this time, the growth of beam size neraly stops (Fig. 4c). This is because that the energy transfer from higher-order modes to the fundamental mode becomes more and more stronger, making this effect based beam size narrowing progressively competes the beam expanding induced by increased mode-energy.

Further increasing the output power to 7.5 W, spectral overlap among these spatial modes and the mode-energy transfer further increase; while at the 10.6 W level, all spectra merge into a single entity, thereby a completely synchronized multicomponent soliton comes into being (SPM broadens the whole spectrum). At this time, the multicomponent soliton acts as a single entity, and therefore can be regarded as an increased mode-area single-mode soliton (similar to most energy converted into the fundamental mode). Strong inter-soliton attraction leads to reduction of the pulse width to ~31.08 ps (Fig. 4a). What important is that the beam size shrinks to 3.72 mm (Fig. 4c) due to more energy concentrated into the fundamental mode, clearly demonstrating the self-focusing like behavior of STSs in multimode fibers. Here, the Kerr nonlinearity helps to balance not only temporal dispersion but also spatial diffraction. This can also be explained as natural thermalization of the multimode beam to increase the amount of entropy (disorder) of the system[30]. It is found that the mode-locking state at ~10 W level is much more stable than that at lower power levels, confirming the significance of strong nonlinear intermodal coupling in the generation and stablization of STSs.

Figure 4 shows the detailed evolution of the multimode soliton pulse profile (autocorrelation curves and sech$^2$ fitting), pulse duration, and laser beam diameter at different output power levels. The pulse width first narrows due to soliton attraction, and then expands due to increased mode number and instability, and finally narrows again because of the formation of single-mode-like soliton (completely synchronized multimode soliton) (Fig. 4b). The laser beam size first increases due to strengthening of higher-order modes, and then narrows due to self-focusing-like energy concentration (Fig. 4c). At the maximum power level, the soliton has the narrowest duration and the smallest beam size (but the highest energy), showing energy condensation of multimode solitons in fiber lasers, which is schematically indicated in Fig. 4d. It is the strong spatiotemporal coupling of the multimode soliton that gives rise to self adjustment of its temporal shape and spatial size. This self organization behavior is highly gain-mediated, which is substantially different from the situation in a passive fiber[12].

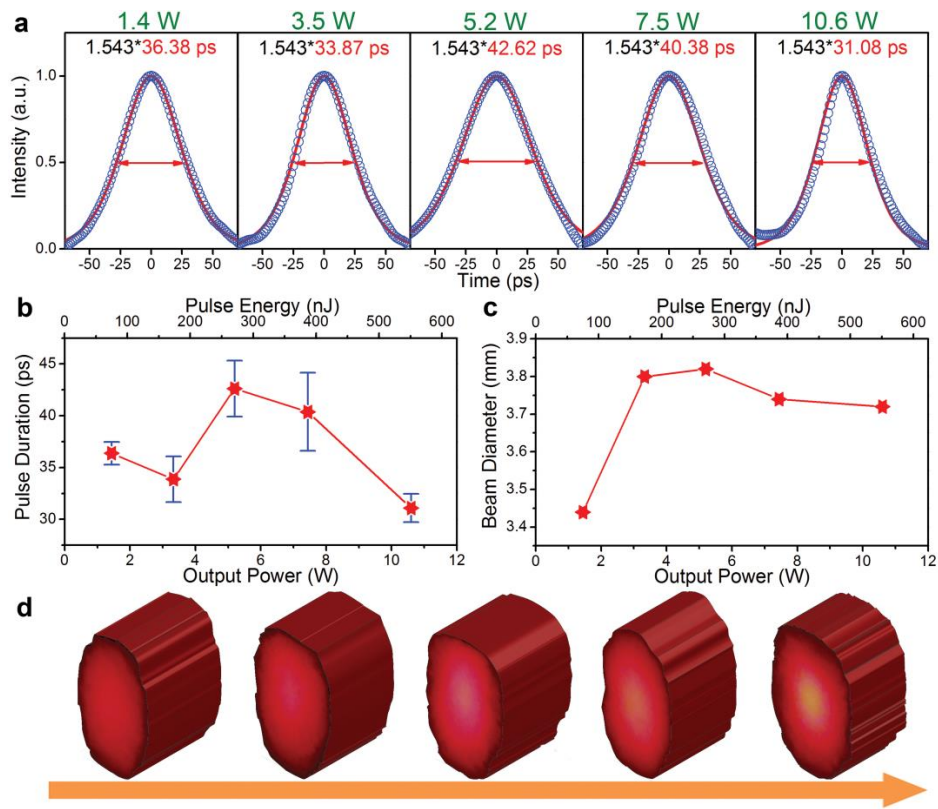

**Figure 4. The multimode soliton pulse shape (a), pulse duration (b) and laser beam diameter (c) of the mode-locked multimode fiber laser at different output power levels. (d) schematically shows the light bullets with the cross section corresponding to the beam size while the longitudinal length corresponding to the transmission distance of each pulse.** In (a), the pulse profile is the measured autocorrelation curve and fitting with the sech$^2$ function. In (c), the beam diameter is measured at 16 mm away from the output fiber facet. In (d), the bottom arrow shows the power-increasing direction.

Figure 5 shows the evolution of laser peak power, power density and energy density of the multimode soliton. All these three properties increase near exponentially with output power, and the maximum peak power is around 17.7 kW. The pulse power (energy) density, which is measured/calculated within the fiber core (cross section and volume of the fiber core are used), can approach 3.6 GW/cm$^2$ (120.4 mJ/cm$^3$). Provided the damage threshold of the mode-locker is increased (e.g., expanding the beam size before impinging on the SESAM), higher pulse energy, peak power, and pulse power (energy) density are expected.

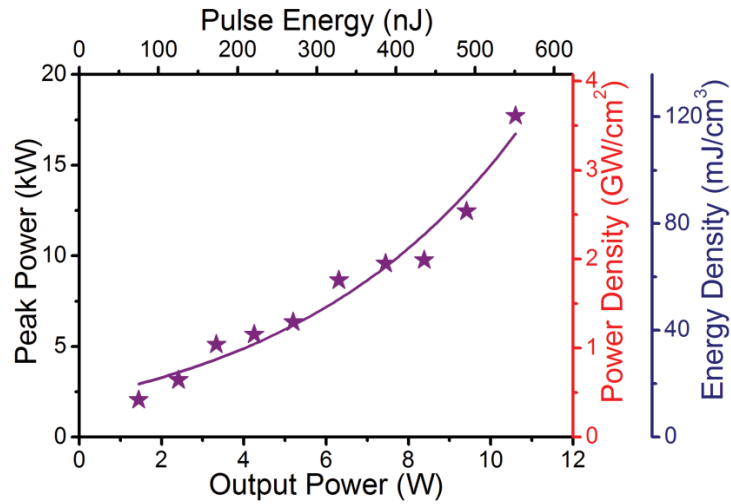

**Figure 5. Output peak power (left vertical axis), power density and energy density (right vertical axes) of the STSs from the mode-locked multimode fiber laser at different output power/energy levels.** The stars are measured and calculated data, while the solid line is exponential fitting.

**Discussion**

As proposed, anomalous dispersion is helpful for the generation of light bullets in bulk media. Therefore, we choose the thulium fiber to carry out the study due to that its 2 μm emission wavelength lies far in the anomalous dispersion regime, which is very beneficial for balancing large modal dispersion of the multimode fiber. Based on calculation from the material parameters (refer to Methods), the modal group velocity delay between the fundamental mode and the 3$^{nd}$ order mode (LP$_{01}$ and LP$_{21}$/LP$_{02}$) is about 37 ps. On the contrary, the material dispersion delay between the fundamental mode and the 3$^{nd}$ order mode is calculated to be ~-5 ps. Therefore, the material dispersion can effectively cancel part of the modal dispersion, and 2 μm fibers provide an appropriate platform to display the formation and dynamics of light bullets.

Transition from conventional single-mode solitons to multimode solitons in step-index multimode fibers is nontrivial due to large modal dispersion and the requirement of phase-locking many different spatial modes simultaneously. In conventional single-mode solitons, Kerr nonlinearity (SPM) is only required to balance the temporal material dispersion. However, this single balance is not sufficient when the spatial degree of freedom is included, where strong modal dispersion can easily lead to pulse splitting. To mode-lock different spatial modes, Kerr nonlinearity must first balance the material dispersion (chromatic dispersion) and then help the material dispersion to exactly cancel the spatial modal dispersion. In that case, the multiple modes are coupled in both the spatial and temporal regimes. In our system, the introduction of gain efficiently mediates the spatiotemporal coupling, by which multiple modes can be successfully phase-locked. The gain-based spatiotemporal coupling (through gain competition and energy redistribution among different modes) also helps to stabilize the multicomponent soliton through self-adjusting its spatial and temporal profiles in a varying circumstance (such as changing the power level). Under high power levels (high light intensity), Kerr nonlinearity is strong enough to balance the spatial diffraction and consequently STSs or light bullets are observed in multimode fibers, which is reminiscent of the self-focusing phenomenon in bulk media[31].

Here, the spatiotemporal coupling of the multimode solitons can bring many unique nonlinear dynamics significantly distinguished from traditional single-mode solitons because the spatial and temporal features of the solitons are highly related. For example, the pulse narrowing will increase the peak light intensity, which will in turn influence the pulse beam size through changing the nonlinearity. Some distinctive nonlinear effects have been observed in graded-index fibers by F. Wise *et al.*, such as multimode four-wave mixing, supercontinuum generation, multiple filamentation, etc.[13] What is different in our system is that the spatiotemporal coupling is mediated by gain, which makes the observed spatial and temporal nonlinear dynamics highly gain dependent, manifesting as pulse narrowing/broadening and beam focusing/expanding under different power levels. Therefore, the incorporation of gain to multimode solitons will not only bring them a self-arrangement capability but will also further enrich the related nonlinear dynamics. This also makes the multimode soliton dynamics highly controllable.

To achieve ultrashort laser pulses in this multimode fiber system (e.g. femtosecond pulses), multimode dispersion compensating fibers or space elements are required to complete the laser cavity. Based on the total spectral width of 8 nm under the maximum pulse energy (Fig. 3e), the compressible Fourier transform-limited pulse width is within 900 fs. However, in practice it is not easy to compensate all these transverse modes simultaneously just with a commonly available dispersion-compensating element due to significantly different dispersion features of each mode. Efficient compensation requires considering both the temporal dispersion and mode dispersion, which will be our future work.

Certainly, with femtosecond pulse duration, the pulse's peak power will be greatly improved; modal coupling and interaction will be stronger, which will lead to new exciting phenomena that require further exploring.

The observation of high-power/energy STSs, soliton attraction, soliton volume squeezing, and inter-mode energy transfer in multimode step-index fibers has great implications in fundamental science and practical applications. Understanding the complicated interaction process and the spatiotemporal coupling of multimode optical solitons will help to uncover the mechanisms of various soliton dynamics and a number of unique phenomena of multidimensional solitary waves. For practice, STSs are valuable in multichannel telecommunications, space-division multiplexing, and high power/energy laser pulse sources. Other potential applications include spectral splicing for pulse compression and multimodal coherent beam combination. Therefore, the observation of high power STSs and related dynamics in multimode fibers will open a new avenue for both solitary waves and multi-dimensional nonlinear optical effects.

## Methods

**Experimental setup**. To conveniently tune the mode-locking state and generate high power laser pulses, a linear laser cavity modulated by a commercial semiconductor saturable absorber mirror (SESAM, BATOP, SAM-2000-43) is adopted, as shown in Fig. 1. This SESAM has high reflection from 1900 nm to 2050 nm, a modulation depth of ~25%, and a saturation fluence of ~35 uJ/cm$^2$. The pump source is a 793 nm high power laser diode (LD) with a maximum output power of 40 W and a 220 μm (0.22 NA) pigtail fiber, which matches to the pump fiber of a (6 + 1) × 1 high power fiber combiner (ITF, MMC0611C4090). The pump light is launched into the gain fiber through this fiber combiner with a coupling efficiency of ~90%. The gain fiber used is a double-clad thulium-doped fiber (TDF, Nufern, PLMA-TDF-25P/400-HE) with a 25 μm diameter, 0.09 numerical aperture (NA) core doped with $Tm^{3+}$ of ~4wt.% concentration. The pure-silica inner cladding, coated with a low-index polymer, has a 400 μm diameter and a NA of 0.46. The absorption of this gain fiber at 793 nm with cladding-pump was measured to be ~3 dB/m with the cut-back method. A ~3.5 m TDF used here is wrapped on a water-cooled copper drum with a diameter of 25 cm, and the fusion splices are clapped in copper heat sinks for efficient cooling. The LD and fiber combiner are also assembled on a water-cooled plate for efficient heat dissipation. The energy-guide fiber, which is used to complete the cavity, has matched parameters to the TDF and a length of 1.6 m. The dispersion of the TDF (energy-guide fiber) at 2 μm is -88 ps$^2$/km (-70 ps$^2$/km), giving a total net dispersion of the cavity of ~0.42 ps$^2$. On the output end, the perpendicularly cleaved fiber facet (~4% Fresnel reflection) is employed to provide laser feedback and acts as the output coupler. In this way, most of the intracavity energy can be extracted. The other end of the cavity fiber is directly butt coupled to the SESAM.

**Experimental measurements**. The laser output power is measured with a power meter (Thorlabs, PM320E) and the laser spectrum is recorded with a mid-infrared spectrum analyzer (SandHouse, SIR5000) with resolution of 0.15 nm. A digital storage oscilloscope (Agilent Technologies, DSO9254A, 2.5GHz) with a high speed photodetector (Newport, 818-BB-51, 12.5GHz), and an autocorrelator (A.P.E., pulseCheck USB MIR) are used to characterize the laser pulsing dynamics. The RF spectra are measured with a RF spectrometer (Agilent, E4402B, 9kHz-3.0GHz) with a resolution of 1 kHz.

**Laser beam shape and size characterization.** The laser beam shape is first displayed on a detector card (Thorlabs, VRC2), which is 66 mm away from the output fiber end, and then recorded with a camera (Cannon, 650D). And the laser beam size is measured at 16 mm away from the output fiber end with the variable round aperture method. The cross section and the longitudinal length of the light bullets (Fig. 4d) correspond to the laser beam size and the transmission distance of each pulse, respectively.

**Acknowledgements**

The authors acknowledge experimental help from Zhipeng Qin and Lingchen Kong, and image processing assistance from Yiren Wang. This work was supported by the National Natural Science Foundation of China under Grant No. 61138006, No. 61275136, and No. 11121504, and the Research Fund for the Doctoral Program of Higher Education of China (No.20120073120085).

**Author Contributions**

Y.W. and Y.T. designed the experiment and wrote the manuscript. Y.W. conducted the experiments. S.Y did the simulation of the mode profile. Y.T. and J.X. supervised the project and revised the manuscript. All authors discussed the results and commented on the manuscript.

**Additional information**

**Competing financial interests:** The authors declare no competing financial interests.

**Supplementary information** accompanies this paper at ***

# Supplementary Information

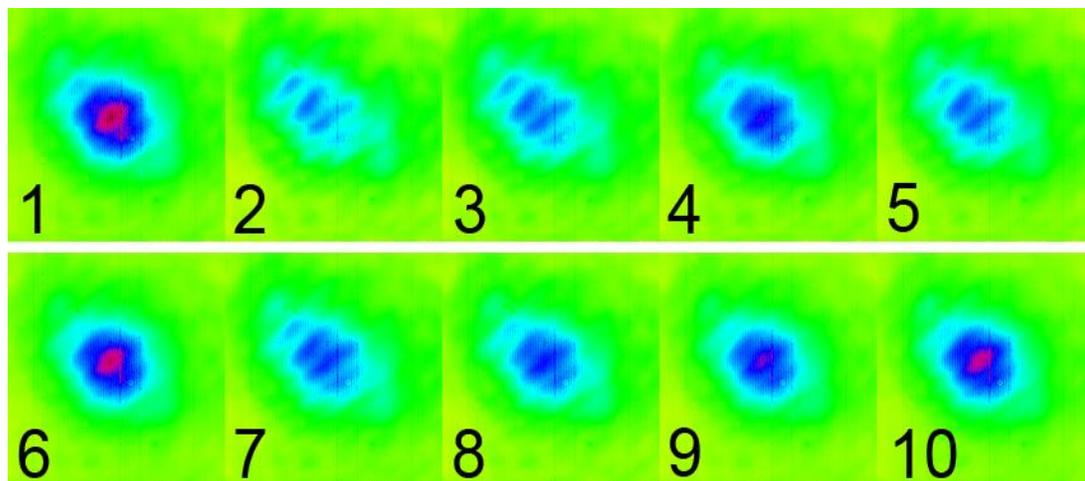

**Supplementary Figure S1.** Laser spot profile measured successively (with 120 ms time increment) of the multimode solitons at the output level of 1.4 W.

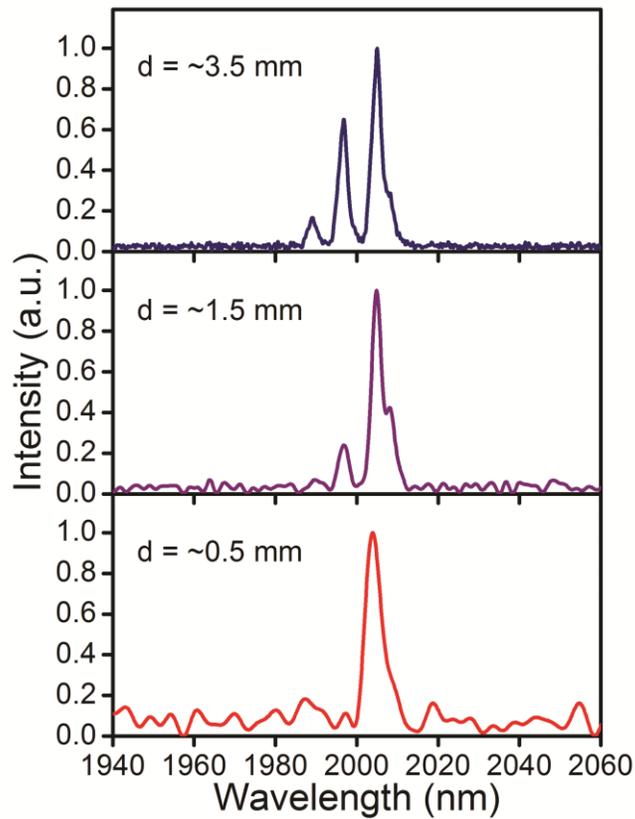

**Supplementary Figure S2. Laser spectra of the mode-locked multimode fiber laser under output power of 1.4 W with a variable round aperture (d is the diameter of the aperture).** The collimated laser beam is coupled into a mid-infrared spectrum analyzer through a variable round aperture. Different spectra are measured with different diameters of the round aperture (~0.5 mm, ~1.5 mm, ~3.5 mm). As it is known that different transverse modes have different mode areas, only the fundamental mode (LP01) with the smallest mode area can be coupled into the spectrum analyzer through the smallest aperture (~0.5 mm) and be detected. So the longest wavelength corresponds to the fundamental mode and other short wavelengths correspond to high-order modes.

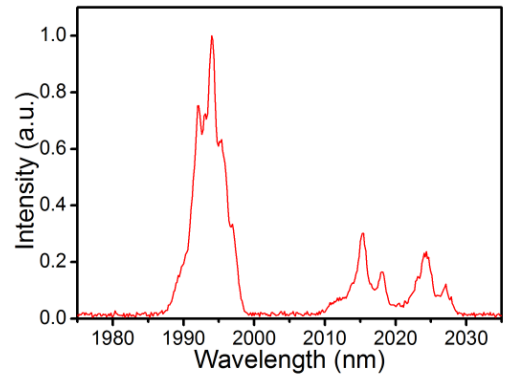
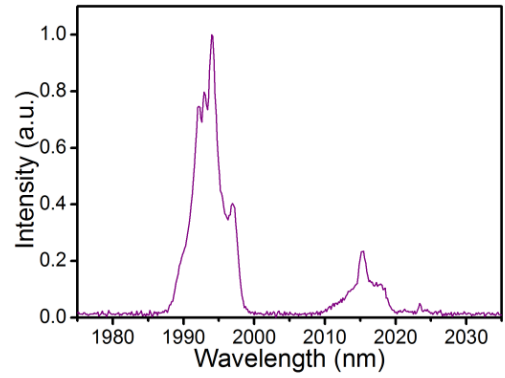
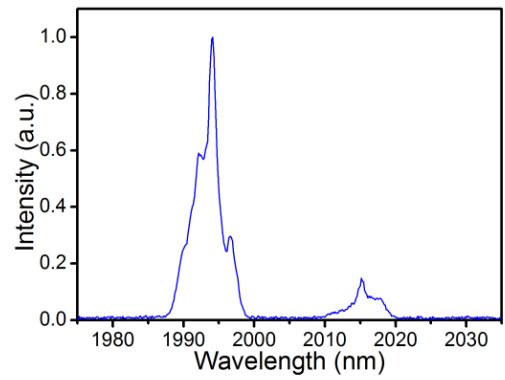

**Supplementary Figure S3. Laser spectra of the mode-locked multimode fiber laser under output power of 1.4 W and at different instants.** The main spectral peaks clearly show the Kelly sidebands.

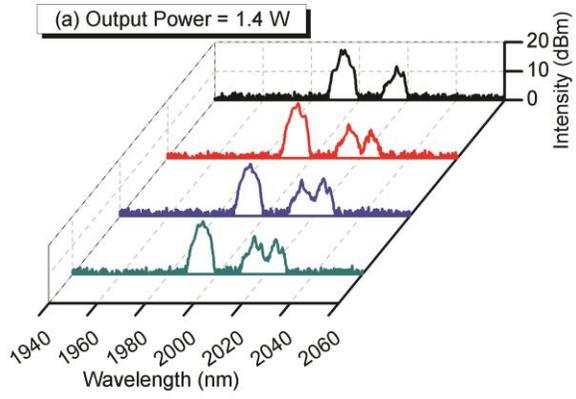
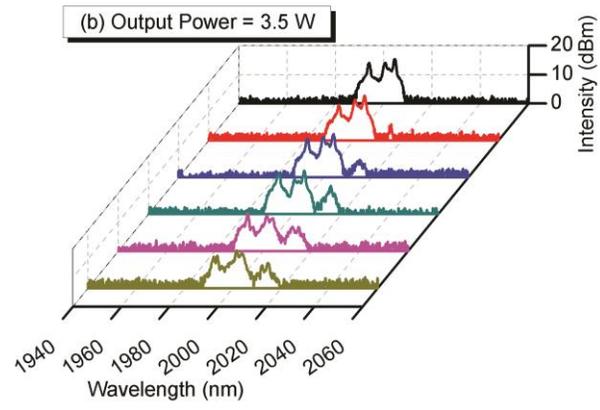
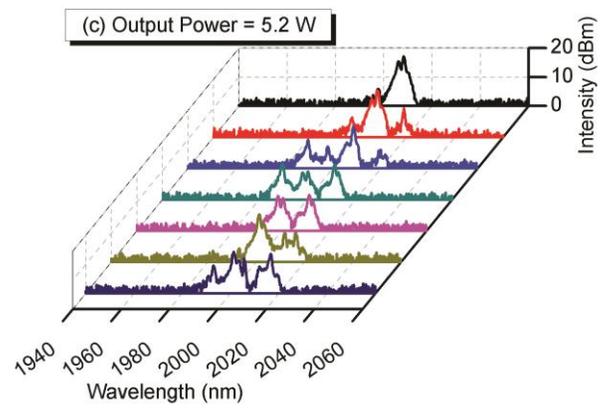

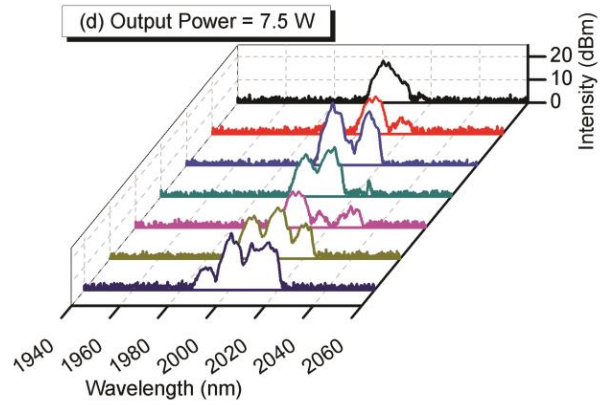

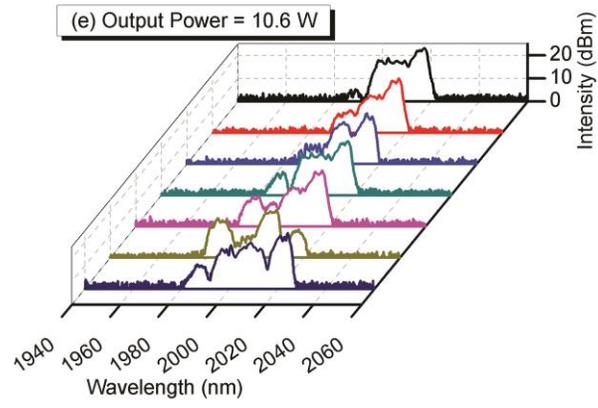

**Supplementary Figure S4. Laser spectra of the mode-locked multimode fiber laser under different output power levels (1.4 W, 3.5 W, 5.2 W, 7.5 W, 10.6 W).** Under each given power level, multiple separately and successively measured spectra are presented.